\documentclass[aps,showpacs,twocolumn]{revtex4}
\usepackage{amsmath}
\usepackage{graphicx}
\usepackage{amsfonts}
\usepackage{amssymb}%
\begin{document}
\title{Is Soret Equilibrium a Non-Equilibrium Effect?}
\author{Alois W\"{u}rger}
\affiliation{Laboratoire Ondes et Mati\`{e}re d'Aquitaine, Universit\'{e} de Bordeaux \&
CNRS, 351 cours de la Lib\'{e}ration, 33405 Talence, France}

\begin{abstract}
Recent thermophoretic experiments on colloidal suspensions revived an old
debate, namely whether the Soret effect is properly described by
thermostatics, or necessarily requires non-equilibrium thermodynamics. Based
on colloidal transport theory and the entropy production of the related
viscous flow, our analysis leads to the conclusion that the equilibrium
approach may work for small ions, yet fails for colloidal particles and
polymers. Regarding binary molecular mixtures, our results shed some doubt on
the validity of thermostatic approaches that derive the Soret coefficient from
equilibrium potentials.

\end{abstract}
\maketitle

\section{Introduction}

Since the early days of thermodynamics there has been a debate whether or not
the stationary state of a non-isothermal system can be described in terms of
equilibrium properties \cite{Gro62}. Classical examples are the Soret and
Seebeck-Peltier effects \cite{Tuc50}: The latter accounts for the coupling
between heat flow and electric current in a conducting material. Though the
Seebeck and Peltier coefficients $S$ and $\Pi$ describe dissipative phenomena,
Thomson showed that irreversibility drops out in the ratio $\Pi/S=T$, which is
simply given by temperature. Several attempts at a rigorous derivation failed,
until Onsager pointed out the role of microscopic reversibility and
established Thomson's relations as special cases of his reciprocal laws
\cite{Ons31}.

The Soret effect describes the mass flow induced by a temperature gradient in
a complex fluid \cite{Wie02,Wie04,Pla06,Pia08a,Wue10}.\ It was first observed
for electrolyte solutions, where salt accumulates at the cold side
\cite{Lud59,Sor79,Chi26}; the non-uniform steady-state concentration $c$\ is
given by the \textquotedblleft Soret equilibrium\textquotedblright%
\ \cite{Eas26,Gro45,Tyr54}
\begin{equation}
\nabla c+cS_{T}\nabla T=0. \label{2}%
\end{equation}
This effect is essential for understanding the compositional grading in the
Earth's petroleum reservoirs \cite{Gal09}, the isotope fractionation in
silicate melts \cite{Hua10}, and the energy balance of carbon-nanotube based
thermogalvanic cells \cite{Hu10}. In microchannels and thin films, the Soret
effect is an efficient means for colloidal confinement and for accumulating
molecular solutes at a micron-sized heated spot \cite{Duh06,Wei08,Jia09,Leo09}%
. In spite of various attempts to elucidate its physical mechanisms, there is
so far no generally accepted theoretical framework for the Soret coefficient
$S_{T}$. A most fundamental question is whether it can be obtained from
equilibrium theory, or whether it necessarily reflects the irreversible nature
of the underlying dissipative flows. This is not a merely formal issue but
affects measurable quantities, e.g., how $S_{T}$ depends on the particle size
\cite{Duh06,Bra08}.

In an early approach Eastman considered two small but macroscopic cells at
temperatures $T$ and $T+dT$ \cite{Eas26}.\ The probability for a particle
moving from one cell to the other is related to the corresponding change of
entropy, absorbed or released by the surrounding liquid. As an essential step
in his argument, Eastman identified this transfer quantity\ with the canonical
entropy $S=-dG/dT$ carried by the solute, where $G$ is the single-particle
free enthalpy or Gibbs energy; his result reads in modern notation
\begin{equation}
S_{T}=\frac{1}{k_{B}T}\frac{dG}{dT}. \label{6}%
\end{equation}
In this thermostatic approach, the Soret coefficient is related to an
equilibrium thermodynamic potential.

A rather different picture emerges from Onsager's non-equilibrium theory, that
is based on the entropy balance equation for reversible and irreversible
changes; the former correspond to the entropy transfer and the latter to
dissipation or entropy production. Heat and mass flows are driven by
generalized forces; the current of solute particles
\begin{equation}
J=-D\nabla c-cD_{T}\nabla T \label{8}%
\end{equation}
accounts for Fick diffusion with the Einstein coefficient $D$ and for
thermophoresis with mobility $D_{T}$ \cite{Gro62}. Comparing the steady-state
condition $J=0$ with (\ref{2}) gives the Soret parameter
\begin{equation}
S_{T}=\frac{D_{T}}{D}. \label{10}%
\end{equation}
Since $D$ and $D_{T}$ do not form a pair of reciprocal coefficients, their
ratio is expected to reflect the underlying dissipative motion. In contrast to
this view, Eq. (\ref{6}) relates $S_{T}$ to an equilibrium free enthalpy.

This discrepancy was already noted by de Groot in his 1945 thesis
\cite{Gro45}. In recent years a controversial discussion aroused from
experiments on collodial suspensions that reported a quadratic \cite{Duh06} or
linear \cite{Put07,Bra08,Jia09,Esl12,Vig07} dependence of the Soret
coefficient on the particle size. Either of these findings is supported by a
number of theoretical works which may be loosely classified as
\textquotedblleft equilibrium\textquotedblright%
\ \cite{Ast07,Bri03,Fay05,Dho07,Sem09}\ and \textquotedblleft
dissipative\textquotedblright\ \cite{Ruc81,Mor99,Ras08,Wue08,Wue09} approaches.

The present paper intends to resolve this discrepancy by investigating the
relation between\ Eqs. (\ref{6}) and (\ref{10}). Starting from a general Gibbs
interaction energy and relying on standard colloidal transport theory, we
evaluate the viscous factors occurring in (\ref{8}) and, in particular,
determine under which conditions they drop out in the ratio $D_{T}/D$. As an
unambiguous signature for dissipation, we calculate the steady-state entropy
production of the viscous flow around a solute particle.

\section{Local thermal equilibrium}

We consider a dilute solution in a non-uniform temperature and assume that
mechanical and local thermal equilibrium is established. This assumption has
several important implications \cite{Gro62}: (i) The macroscopic pressure is
constant throughout the system.\ (ii) The Dufour effect being small in
liquids, the temperature profile $T(\mathbf{r})$ is independent of composition
and constant in time. (iii) More generally, the properties of a small but
macroscopic subvolume are described by equilibrium thermodynamics.

Thus we may define a free-enthalpy density $g(r,T(\mathbf{r}))$ that describes
the mutual forces of a particle and the surrounding fluid. It depends on
position both explicitly through the interaction and implicitly through the
non-uniform temperature. Relevant mechanisms are the electric-double layer
energy, the van der Waals potential, and depletion forces due to an additional
molecular solute. The solvation free enthalpy or Gibbs energy of a single
particle is given by
\begin{equation}
G(T(\mathbf{r}))=\int dVg(r,T(\mathbf{r})). \label{16}%
\end{equation}
The integrand is most significant within the range of interaction $\lambda$
and rapidly decreases at larger distances \cite{And89}. For electric-double
layer forces, $\lambda$ corresponds to the Debye length and $g$ decays
exponentially as $e^{-(r-R)/\lambda}/r$. Dispersion forces have no intrinsic
length scale but decrease with a power law; then $\lambda$ may be taken as the
particle size. The main conclusions of this paper are very general and apply
to any interactions as long as $g(r,T(\mathbf{r}))$\ decays sufficiently
rapidly at large distances.

For low dilution and in the absence of long-range interactions, the chemical
potential consists of the single-particle Gibbs energy and a contribution
accounting for the translational entropy $-k_{B}\ln(c/c_{0})$,
\begin{equation}
\mu(T,c)=G(T)+k_{B}T\ln(c/c_{0}). \label{17}%
\end{equation}
The spatially varying temperature $T(\mathbf{r})$ is imposed by the
experimental setup, whereas the concentration profile $c(\mathbf{r})$ remains
to be determined.

\section{Thermostatic approach}

We briefly discuss the origin of the relation (\ref{6}), which was first
obtained by Eastman for the Soret effect of electrolyte solutions and since
then has been applied to molecular mixtures and colloidal particles. Eastman
explicitly discards dissipative processes and retains reversible changes of
the thermodynamic potential only. Considering a\ particle that migrates
between \textquotedblleft cells\textquotedblright\ with temperature and
concentration differences $dT$ and $dc$, he defines the \textquotedblleft
entropy of transfer\textquotedblright\ $S^{\ast}$ through the change of the
chemical potential due to the uniform concentration
\[
S^{\ast}dT=-(d\mu/dc)dc.
\]
(At this point, the unknown Soret coefficient $S_{T}$ has merely been replaced
by the unknown $S^{\ast}$.) As the essential step of his approach, Eastman
then identifies $S^{\ast}$ with the canonical single-particle entropy $S$ and
thus obtains (\ref{6}).\ 

Subsequent works derived Eastman's formula from the condition that the
gradient of the free energy or the chemical potential vanishes. The underlying
idea is to interprete $\nabla\mu$ as the mechanical force acting on a particle
and, accordingly, the relation $\nabla\mu=0$ as the steady-state condition. To
linear order in the gradients one has
\begin{equation}
0=\nabla\mu=\frac{dG}{dT}\nabla T+k_{B}T\frac{\nabla c}{c}. \label{30}%
\end{equation}
Comparing with (\ref{2}) gives indeed Eastman's expression for the Soret
coefficient. (In neglecting the term proportional to $\ln(c/c_{0})\nabla T$
one circumvents a problem arising from the fact that $\nabla\mu$ is not
invariant under a shift of the zero of entropy. For a discussion of
alternative choices see Ref. \cite{Bri11}.)

This thermostatic approach relies on identifying $\nabla\mu=0$ with the
condition of mechanical equlibrium which is central to the steady state of
dissipative processes \cite{Gro62}. Strictly speaking, the Gibbs energy and
the chemical potential are not well defined for a system with non-uniform
temperature; thus the steady-state condition Eq. (\ref{30}) is beyond the
domain of equilibrium statistical mechanics. Eastman's approach heavily relies
on identifying $S^{\ast}$ with the canonical entropy $S$; so far there is no
justification for this essential step in his argument.

\section{Transport coefficients and size dependence}

Here we evaluate the Soret coefficient from Onsager's theory for
non-equlibrium systems. For small particles we start from the generalized
thermodynamic force. For large particles, the thermodiffusion and diffusion
currents have to be evaluated separately because of their different viscous properties.

\subsection{Diffusion current}

Brownian motion of a particle suspended in a liquid was related by Einstein to
the thermal agitation of nearby molecules, which acts as a random external
force. There are two important consequences, the mean-square displacement
increases linearly with time, and a concentration gradient results in a
diffusion current $-D\nabla c$, where the coefficient is given by the
Stokes-Einstein relation
\begin{equation}
D=\frac{k_{B}T}{6\pi\eta R}. \label{20}%
\end{equation}
This expression is valid over a large range in solute size, from small
molecules of a few \AA \ to large colloidal particles that are visible to the
naked eye. The same law holds true for diffusion of polymers, with $R$
corresponding to the gyration radius \cite{deG79}. For later use we give the
velocity corresponding to gradient diffusion,
\begin{equation}
u_{D}=-D\frac{\nabla c}{c}. \label{20a}%
\end{equation}%

\begin{figure}[b]
\includegraphics[width=\columnwidth]{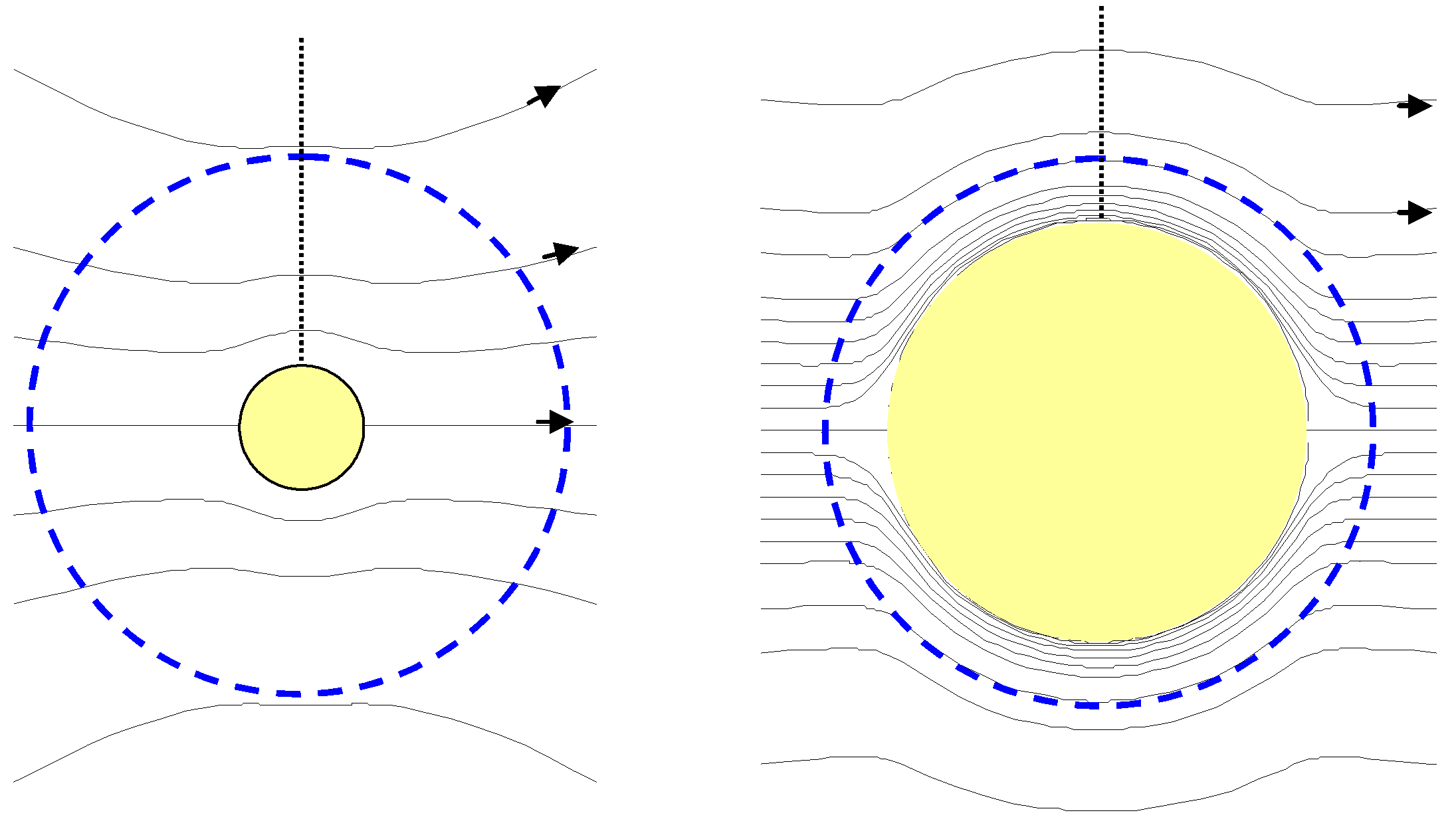}
\caption{Flow pattern $\mathbf{v}(\mathbf{r})$ in the vicinity of an immobile
solute particle. The velocity is zero at the particle surface, increases
exponentially within the interaction length $\lambda$ indicated by a dashed
circle, and decays as $1/r$ at larger distances The left and right panels show
the cases of small and large particles, respectively. The velocity profile
along the vertical dotted lines is shown in Fig. 2.}%
\label{Fig4}%
\end{figure}

\subsection{Thermodiffusion of small particles $R\ll\lambda$.}

In the framework of non-equilibrium thermodynamics, the generalized force
exerted on a dispersed particle reads $\nabla(\mu/T)$ \cite{Gro62}. Inserting
(\ref{17}) and multiplying with $cT$, we find
\begin{equation}
cT\nabla\frac{\mu}{T}=cT\nabla\frac{G}{T}+k_{B}T\nabla c=-c\frac{H}{T}\nabla
T+k_{B}T\nabla c. \label{20b}%
\end{equation}
We have used the implicit spatial variation of the free enthalpy
$G(T(\mathbf{r}))$ and, in the second equality, the Gibbs-Helmholtz equation
$d(G/T)dT=-H/T^{2}$, which epresses the relation $H=G+TS$ between free
enthalpy $G$, enthalpy $H$, and entropy $S$.

The Soret coefficient is obtained from two conditions. First, we identify the
stationary state with zero thermodynamic force, $\nabla(\mu/T)=0$. Second, we
assume that the mobility parameter of both force terms in (\ref{20b}) is given
by Stokes friction $1/(6\pi\eta R)$. This is obvious for the diffusion term
(\ref{20}) and provides a good approximation for the thermodiffusion current
in the limit $R\ll\lambda$, in analogy to H\"{u}ckel's treatment of
electrophoresis \cite{Hie97,Mor08}. Comparison with (\ref{8}) then gives
\begin{equation}
D_{T}=-\frac{1}{6\pi\eta R}\frac{H}{T}.\label{22}%
\end{equation}
Corrections to this approximate result are of the order $R/\lambda$. 

The Soret coefficient is given by the ratio
\begin{equation}
\frac{D_{T}}{D}=\frac{d}{dT}\left(  \frac{G}{k_{B}T}\right)  =-\frac{H}%
{k_{B}T^{2}}.\label{18}%
\end{equation}
Irreversibility drops out since $D$\ et $D_{T}$\ carry the same friction
coefficient.\ Eq. (\ref{18}) states that for particles much smaller than the
range of the solute-solvent interaction, the Soret coefficient is given by the
solvation enthalpy $H$. A particle that strongly interacts with the solvent
$(H<0)$ moves to the cold, whereas a solute with a positive solvation enthalpy
is driven to the warm.

The above result significantly differs from Eastman's formula. The latter
provides a good approximation only if $G/T$ is small as compared to the
derivative $dG/dT$. This seems to be the case for small ions in an electrolyte
solution: The Soret coefficient obtained by Dhont \cite{Dho07} from (\ref{6})
corresponds to the result from the non-equilibrium thermodynamics approach
\cite{Mor08}. The thermodynamic force (\ref{20b}) agrees with the Boltzmann
type distribution function $c=c_{0}e^{-G/k_{B}T_{0}}$ that was used by Duhr
and Braun \cite{Duh06} and confirmed by Astumian \cite{Ast07}. The general 1D
steady-state distribution was given by van Kampen in a study on diffusion in
non-uniform media \cite{Kam88}. As a common feature these works assume, more
or less explicitly, that the dissipative factor of the drift velocity $u_{0}$
is given by Stokes-Einstein form $6\pi\eta R$; this is justified for small
particles of radius $R\ll\lambda$.

\subsection{Thermophoresis of large particles $R\gg\lambda$.}

Now we turn to the opposite case illustrated in the right panel of Fig. 1.
Then the viscous stress on the fluid is concentrated in a boundary layer of
thickness $\lambda$ at the particle surface, where Stokes' equation takes a
rather simple form \cite{And89}. From a general argument relying on the
symmetry of Onsager's coefficients, Derjaguin calculated the quasi-slip
velocity \cite{Der87} which gives the thermophoretic mobility in the form
\begin{equation}
D_{T}=-\frac{2}{3\eta}\frac{\hat{h}}{T}, \label{24}%
\end{equation}
where an integral over the solvation enthalpy density $h(z,T)$,%
\begin{equation}
\hat{h}=\int_{0}^{\infty}dzzh(z,T).
\end{equation}
In the latter integral the quantity $z=r-R$ is the distance from the particle surface.

For large particles both the enthalpy density $h$ and the quantity $\hat
{h}\sim\lambda^{2}g(0)$ are independent of the radius $R$, whereas the volume
integral $H=4\pi R^{2}\int dzh$ is proportional to the surface area. As a most
important feature, the velocity $u_{0}=D_{T}\nabla T$ and the mobility $D_{T}$
are independent of the solute size; this is a particular case of a general
rule for motion driven by interfacial forces \cite{And89}. The above form for
the diffusion coefficient implies that the ratio%
\begin{equation}
\frac{D_{T}}{D}=\frac{4\pi R}{k_{B}T}\frac{\hat{h}}{T}\propto R \label{28}%
\end{equation}
increases linearly with\ the particle radius. This result generalizes previous
work on electrostatic and dispersion forces
\cite{Ruc81,Mor99,Ras08,Wue08,Wue09}; for charged particles, the double-layer
enthalpy $h$ comprises the energy density of the electric field and the excess
ion osmotic pressure \cite{Wue08}. The underlying hydrodynamic treatment of
interfacial forces parallels Smoluchowski's treatment of electrophoresis and
is widely used in colloidal transport theory \cite{And89}. We have discarded
the possiblility of hydrodynamicc slippage \cite{Mor09}.

\subsection{Polymers}

Regarding the ratio (\ref{10}) we first note that the diffusion coefficient of
polymers is proportional to the inverse gyration radius $D=k_{B}T/\kappa\eta
R_{g}$, due to long-range hydrodynamic interactions between the flexible units
of the polymer \cite{deG79}. The thermophoretic mobility $D_{T}$ is
independent of the molecular weight, as first derived by Brochard and de
Gennes from general arguments; for more detail see \cite{Wue07,Wue10}. In
physical terms, the Gibbs energy $g$ does not cause hydrodynamic interactions
beyond the interaction range $\lambda$. Thus Eq. (\ref{28}) remains valid for
polymers. Since the gyration radius scales with the number $N$ of monomers
according to the power law $R_{g}=\ell N^{\nu}$, the Soret coefficient%
\begin{equation}
\frac{D_{T}}{D}\propto R_{g} \label{7a}%
\end{equation}
shows a characteristic molecular-weight dependence. As a side remark, this
scaling ceases to be valid for chains of less than hundred monomers, where an
additional thermodiffusion mechanism sets in \cite{Sta09,Wue09,Kit12}.

\subsection{Comparison with experiment}

\textit{Polymers. }As first shown by Giddings and co-workers \cite{Gid76}, the
Soret coefficient of high polymers is proportional to the gyration radius
$R_{g}$, ,
\begin{equation}
S_{T}\propto R_{g}\sim\ell N^{\nu}\ \ \ \ \ \ \ \ \text{(exp)} \label{7}%
\end{equation}
For ideal flexible polymers, $\ell$ is diameter of a monomer and $\nu=\frac
{1}{2}$. For most real polymers, $\ell$ rather corresponds to the persistence
length and is larger than a monomer; accordingly, $N$ gives the number of such
units. The exponent depends on intrachain interactions and the solvation
energy; in a \textquotedblleft good solvent\textquotedblright\ one has
$\nu\approx0.6$. The experimental law is in perfect agreement with the
expression obtained from the ratio of transport coefficients (\ref{7a}).

For sufficiently long chains the Gibbs energy is proportional to the molecular
weight, $G=Ng_{1}$, and so is Eastman's expression,%
\begin{equation}
S_{T}=Ns_{1}\text{,}%
\end{equation}
with the monomer Soret coefficient $s_{1}=-(k_{B}T)^{-1}h_{1}/T$. This linear
dependence does not match the experimental finding. Since the scaling law
$H=Nh_{1}$ is a fundamental thermodynamic property of high polymers, the
failure of Eastman's expression can not be mended. The same argument holds
true for similar approaches based on the chemical potential, which for a long
chain is an extensive quantity, $\mu=N\mu_{1}$.

\textit{Colloidal particles. }\ We discuss the size dependence of the Soret
coefficient of large colloidal particles. For charged polystyrene beads
confined in an 10-micron chamber, a quadratic variation $S_{T}\propto R^{2}$
was found for radii ranging from 20 nm to 1 $\mu$m \cite{Duh06}, whereas
subsequent experiments reported a linear dependence $S_{T}\propto R$ for both
solid particles \cite{Put07,Bra08,Jia09,Esl12} and microemulsion droplets
\cite{Vig07} \ 

The Gibbs energy of large particles is proportional to the surface area,
$G\propto R^{2}$, resulting in the quadratic law $S_{T}\propto R^{2}$ for
Eastman's expression (\ref{6}), which has been used for fitting the data of
Ref. \cite{Duh06}. On the other hand, since the Einstein coefficient is
inversely proportional to the radius and $D_{T}$ independent of the particle
size \cite{And89,Mor99,Wue10}, the ratio (\ref{28}) results in $S_{T}\propto
R$, in agreement with the data of Refs. \cite{Put07,Bra08,Jia09,Esl12,Vig07}.
A more complex situation occurs for strongly hydrophobic particles with a
finite slip length, where a quadratic dependence is obtained for intermediate
particle size \cite{Mor08}; yet this effect can be discarded for micron-sized beads.

\section{Entropy production}

Here we relate the title of this paper to the question whether or not the
Soret equilibrium (\ref{2}) produces entropy. The thermostatic approach
(\ref{6}) relies on the assumption that the non-uniform solute distribution
does not contribute to dissipation. This is obvious from Eastman's argument:
The \textquotedblleft entropy of transfer\textquotedblright\ $S=-dG/dT$
accounts only for the reversible change that occurs while a particle migrates
between regions of different temperatures. In this picture, the non-uniform
steady state described by (\ref{2}) is not related to dissipation and, as a
consequence, does not produce entropy.\ 

\subsection{Relevant dissipation mechanisms}

\begin{figure}
\includegraphics[width=\columnwidth]{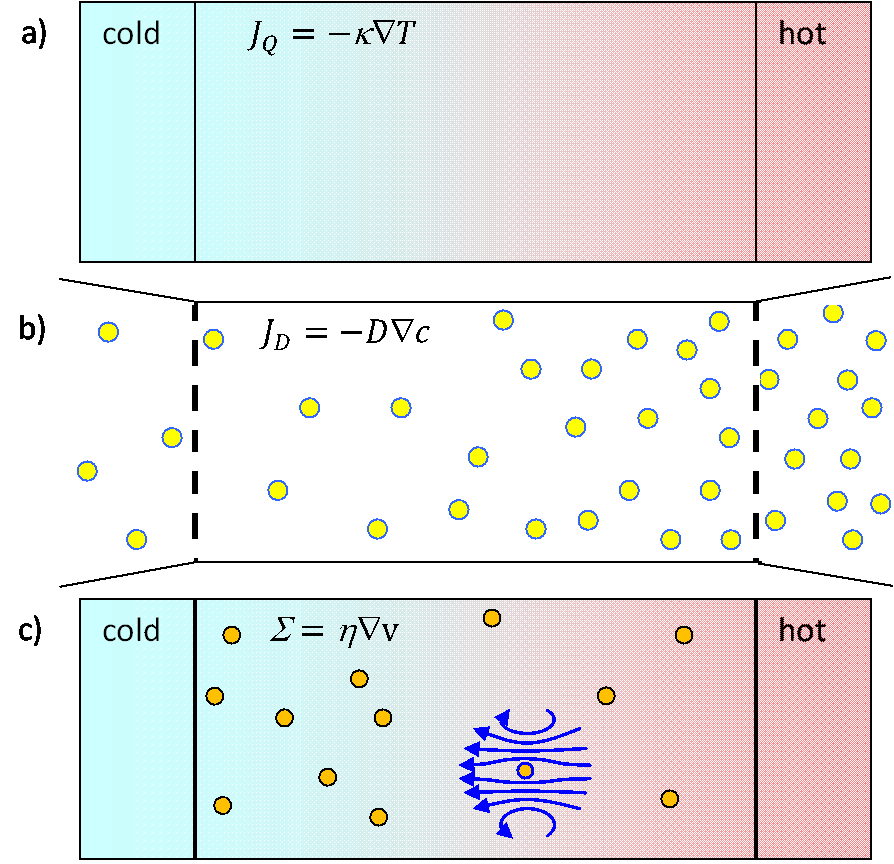}
\caption{Non-equlibrium systems~with stationary heat and matter flow. a) A
constant temperature gradient drives a heat flow $J_{Q}=-\kappa\nabla T$
\ from the hot to the cold side. b) An open system is in contact with two
reservoirs at different solute concentration $c$. Gradient diffusion results
in a steady particle current $J=-D\nabla c$ from high to low concentration,
with Einstein coefficient $D$. c) In the steady state of a closed system, the
particle current vanishes. Yet the viscous flow $\mathbf{v}(\mathbf{r})$ in
the vicinity of each solute particle dissipates energy. These situations
correspond to the three terms of the rate of entropy production (\ref{9a}). }%
\label{Fig2}%
\end{figure}

As a fundamental aspect of Onsager's non-equilibrium thermodynamics,
dissipation is related to flows of heat and matter. Here we consider three
contributions to the rate of entropy production \cite{Gro62},
\begin{equation}
\sigma=-\frac{J_{Q}\cdot\nabla T}{T^{2}}-\frac{J\cdot k_{B}\nabla c}{c}%
-\frac{\Sigma:\mathbf{\nabla v}}{T}. \label{9a}%
\end{equation}
The first one accounts for heat flow $J_{Q}$ from the hot to the cold side of
the sample, as illustrated in Fig. \ref{Fig2}a. The steady-state heat current
$J_{Q}=-\kappa\nabla T$ is proportional to the temperature gradient and the
thermal conductivity $\kappa$.%

The second term in (\ref{9a}) describes dissipation due to a particle flow $J$
in a concentration gradient $\nabla c$. The steady state of a closed system
corresponds to $J=0$. As an example for a finite stationary particle current,
Fig. \ref{Fig2}b shows an open system at constant temperature that is in
contact with two reservoirs at different concentration. This concentration
gradient results in a stationary current $J=-D\nabla c$, which is constant
throughout the sample. Since the entropy per particle is higher in the
low-concentration reservoir, the entropy per unit time leaving at the left
boundary exceeds that entering at the right. This net outward flow is
supplemented by the entropy production within the system. For later use we
give the entropy production per particle, \
\begin{equation}
\dot{S}_{D}=k_{B}\frac{u_{D}^{2}}{D}=\frac{6\pi\eta u_{D}^{2}}{T}R. \label{12}%
\end{equation}

The last term in $\sigma$ is given by the contraction of the tensor of
velocity derivatives $\mathbf{\nabla v}$ and the symmetric part of the viscous
stress $\Sigma$. In the stationary state the latter reads $\Sigma=-\frac{1}%
{2}\eta(\mathbf{\nabla v+\nabla v}^{\top})$, with the solvent viscosity $\eta$
\cite{Gro62}. The total entropy production $\dot{S}$ due to the viscous flow
in the vicinity of a single particle is obtained by spelling out the
components of the stress tensor $\Sigma$ and taking the volume integral
\cite{Gro62}, \
\begin{equation}
\dot{S}=\frac{\eta}{2T}\sum_{ij}\int dV\left(  \frac{dv_{j}}{dx_{i}}%
+\frac{dv_{i}}{dx_{j}}\right)  ^{2}. \label{21}%
\end{equation}
In the following we show that the solute particles dissipate energy, even in
the steady state $J=0$. Though the entropy production is much smaller than the
heat-current driven term $\kappa(\nabla T)^{2}/T^{2}$ and thus difficult to
measure, it is an important signature for the dissipative nature of the Soret
equilibrium state.

In the following we consider the entropy production due to the solute
particles. The steady state $J=0$ is attained if diffusion and drift
velocities cancel each other, $u_{D}+u_{T}=0$. Each of the contributions to
(\ref{8}) engenders characteristic flow patterns in the vicinity of each
particle, which are denoted $\mathbf{v}_{D}(\mathbf{r})$ and $\mathbf{v}%
_{T}(\mathbf{r})$, respectively. As a most important aspect we discuss the
implications of the steady-state condition on the total velocity field%
\begin{equation}
\mathbf{v}=\mathbf{v}_{D}+\mathbf{v}_{T}. \label{11}%
\end{equation}
Zero mean velocity of the solute particle ($J=0$) requires that the fluid
velocity field vanishes at its surface, $\mathbf{v}|_{R}=0$.\ Yet this does
not imply that $\mathbf{v}$ is zero everywhere. Quite to the contrary, the
well-known long range behavior $\mathbf{v}_{D}\sim1/r$ and $\mathbf{v}_{T}%
\sim1/r^{3}$ implies that $\mathbf{v}$ is finite at distances beyond the
particle radius ($r>R$) \cite{Lon01,And89}. Thus each solute acts as a pump
that stirs the surrounding fluid as shown schematically in \ref{Fig2}c.

The mean velocity field related to gradient diffusion reads
\[
\mathbf{v}_{D}(\mathbf{r})=\frac{R}{2r}\left(  1+\mathbf{\hat{r}\hat{r}%
}\right)  \cdot\mathbf{u}_{D},
\]
where $\mathbf{\hat{r}}=\mathbf{r}/r$ is the radial unit vector and
$\mathbf{u}_{0}$ the velocity at the particle surface. The term arising from
the thermophoretic drift cannot be given in closed form; in the following we
discuss two limiting cases with respect to the particle size.

\subsection{Small particles $R\ll\lambda$}

For the entropy production $\dot{S}$ we need the net velocity field
$\mathbf{v}=\mathbf{v}_{D}+\mathbf{v}_{T}$.\ For an exponentially screened
interaction free enthalpy, one finds to lowest order in $R$ \cite{Lon01,Mor08}%
\[
\mathbf{v}=(e^{-(r-R)/\lambda}-1)\mathbf{v}_{D}.
\]
This means that $\mathbf{v}_{D}$ and $\mathbf{v}_{T}$ cancel each other at the
particle surface.\ This expression constitutes a poor approximation for
$r\gg\lambda$; yet this range is of little interest here.) Calculating the
viscous stress tensor in spherical coordinates and inserting in (\ref{21}),
one obtains
\begin{equation}
\dot{S}=\frac{14\pi\eta u_{D}^{2}}{T}\frac{R^{2}}{\lambda}\ln\frac{\lambda}%
{R}\ . \label{19}%
\end{equation}
Thus in the limit of a point particle $R/\lambda\rightarrow0$, the entropy
production vanishes, and the Soret equilibrium is indeed an equilibrium
effect. In turns out instructive to express $\dot{S}$ through the entropy
production of a diffusion current given in (\ref{12}),
\begin{equation}
\dot{S}=\frac{7}{3}\frac{R}{\lambda}\ln\frac{\lambda}{R}\dot{S}_{D}%
\ \ \ \ \ (R\ll\lambda). \label{19a}%
\end{equation}
The small prefactor implies $\dot{S}\ll\dot{S}_{D}$; in other words the
dissipation related to the Soret equilibrium of small particles is much
smaller than that of a corrersponding diffusion current.

\subsection{Large particles}

Now we turn to the entropy production $\dot{S}$. In the boundary layer the
fluid velocity changes by $u_{T}$, resulting in a shear rate of the order
$u_{T}/\lambda$. Integrating $\sigma=(\eta/2T)(u_{T}/\lambda)^{2}$ over the
interaction volume $4\pi R^{2}\lambda$ and noting $u_{T}=-u_{D}$, we find
\begin{equation}
\dot{S}=\xi\frac{6\pi\eta u_{D}^{2}}{T}\frac{R^{2}}{\lambda}, \label{29}%
\end{equation}
where $\xi$ is a numerical factor of the order of unity that depends on the
precise form of the velocity field. At distances well beyond the interaction
length, the velocity $v\sim u_{T}(R/r)$ results in a shear rate $u_{T}R/r^{2}%
$.\ Its contribution to $\dot{S}$ is of the form $\sim(\eta/T)u_{T}^{2}R$,
which is by a factor $\lambda/R$ smaller than (\ref{29}) and thus may be
neglected. Like in the small particle case, the entropy source strength varies
with the square of the solute size. Inserting that of the diffusion current,
we have
\begin{equation}
\dot{S}=\xi\frac{R}{\lambda}\dot{S}_{D}\ \ \ \ \ \ \ (\lambda\ll R).
\label{29a}%
\end{equation}
Thus the single-particle entropy production of the Soret equilbrium is much
larger than that of the corresponding diffusion current.%

\begin{figure}[b]
\includegraphics[width=\columnwidth]{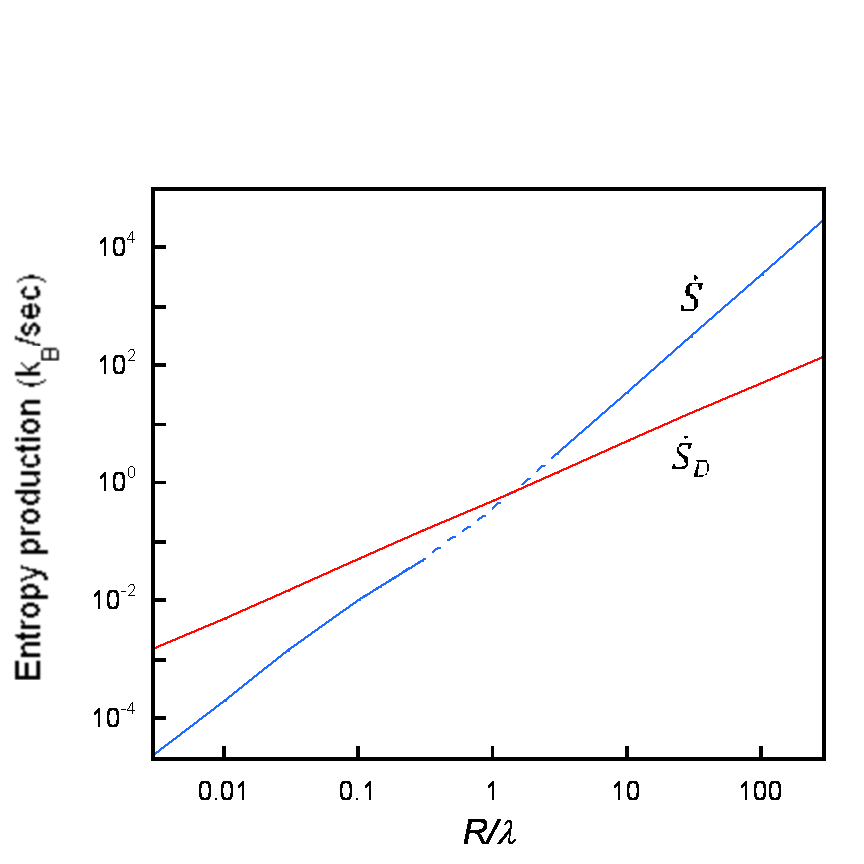}
\caption{Entropy production $\dot{S}$ as a function of the reduced particle
radius $R/\lambda$. The two branches of the curve $\dot{S}$ indicate the
limiting laws (\ref{19}) and (\ref{29}) for the entropy produced by the
viscous flow around a single particle; the dashed line is an interpolation.
The curve $\dot{S}_{D}$ shows the entropy production (\ref{12}) due to a
diffusion current. We have used the parameters $u_{0}=1\mu$m/sec,
$\lambda=100$ nm, the viscosity of water, and the numerical factor $\xi=1$.
The relative magnitude of these two dissipation mechanisms is obvious from
Eqs. (\ref{19a}) and (\ref{29a}). At $R=\lambda$, a cross-over occurs in the
relative dissipation strength of the steady states illustrated in Figs.
\ref{Fig2}b and \ref{Fig2}c: For small particles, maintining a given
concentration gradient through the Soret effect dissipates less energy than
gradient diffusion in an open system, $\dot{S}<\dot{S}_{D}$, whereas for big
particles the Soret equilibrium produces more entropy, $\dot{S}>\dot{S}_{D}$.
}%
\label{Fig6}%
\end{figure}

\subsection{Conclusion}

From the argument given below (\ref{11}) it is clear that the Soret effect of
any solute engenders a finite velocity field $\mathbf{v}$ and thus dissipates
energy at a finite rate $T\dot{S}$ per particle. Strictly speaking, this
implies that the Soret equilibrium is not a true equilibrium property, yet
does not exclude that Eastman's approach describes experimental findings.

We found it instructive to compare the entropy production per particle of the
Soret equilibrium, $\dot{S}$, to that of diffusion with the same concentration
gradient, $\dot{S}_{D}$.\ In the limit $R\ll\lambda$, where the particle
radius is much smaller than the interaction range, the Soret effect dissipates
little energy and, according to (\ref{19a}), produces less entropy than the
corresponding diffusion current. In physical terms, this is related to the
weak fluid flow in the vicinity of the particle and to the partial
cancellation of $\mathbf{v}=\mathbf{v}_{D}+\mathbf{v}_{T}$. As a consequence,
the Soret coefficient is well described by Eastman's thermostatic approach
(\ref{6}), in spite of its non-equilibrium nature.

On the contrary, in the large-particle limit $\lambda\ll R$, the non-uniform
colloidal concentration is a source of significant dissipation. The rate of
entropy production (\ref{29a}) by far exceeds that of the corresponding
diffusion current $\dot{S}_{D}$. This excess dissipation occurs in the
boundary layer within one interaction length from the particle surface, and
arises from the large shear rate that is characteristic for surface forces. In
this range the Soret effect cannot be viewed as an equilibrium phenomenon.

\section{Binary molecular mixtures}

The above laws for polymers and solid particles heavily rely on Stokes
hydrodynamics. Simple results are obtained in the dilute limit for particles
that are much smaller or much larger than their range of interaction. A more
intricate situation is encountered for binary mixtures of non-ionic molecules:
Since the range of dispersion forces is given by the molecular size, there is
no clear separation of length scales; as a consequence, hydrodynamic and
interaction effects occur at similar distances are not easily distinguished.
In general none of the components is dilute.\ Thus the surrounding of a given
molecule consists of all species; the low-dilution limit does not apply, and
thermodiffusion of a given species has to be evaluated self-consistently.\ 

\subsection{Thermostatic approach}

In order to avoid these difficulties, thermodiffusion of non-ionic molecular
liquids has been described in a thermostatic approach, where the Soret
coefficient
\begin{equation}
S_{T}=-\frac{Q_{1}-Q_{2}}{k_{B}T^{2}}%
\end{equation}
is given by the heat $Q_{i}$ carried by each of the components. Their
difference $Q_{1}-Q_{2}=TS$ is related to Eastman's entropy of transfer $S$
introduced above (\ref{6}). Dissipative aspects have been discussed by
identifying $Q$ with Eyring's viscous activation energy $E$ that is defined
through the temperature dependence of the viscosity $\eta=\eta_{0}e^{E/k_{B}%
T}$ \cite{Den52,Dou55,Shu98,Esl09}; a similar picture arises when relating $Q$
to partial enthalpies \cite{Kem01}, partial volumes \cite{Haa50} or the
self-diffusion activation energy \cite{Tic56}. A refined description for the
mutual interactions of the molecular species is achieved by introducing
thermodynamic or \textquotedblleft activity\textquotedblright\ factors in the
chemical potential or the partial pressure \cite{Ber97,Mor09a}.

\subsection{Hydrodynamic approach}

Thermostatic approaches assume, more or less explicitly, that the viscous or
mobility factors of $D$ and $D_{T}$ cancel each other. The above results
suggest that this is not necessarily a good approximation. Since the
interaction range of dispersion forces is comparable to the molecular size,
$\lambda\sim R$, the small-particle limit is not well justified, and one
rather expects that $D$ and $D_{T}$ carry different kinetic or hydrodynamic factors.

Though it may seem questionable at first sight, macroscopic hydrodynamics
works surprisingly well at the molecular scale: The Stokes-Einstein
coefficient (\ref{20}) provides a good description for the diffusion of small
molecules and even of ions. Even an additional coarse-graining in mesoscale
simulations does not affect colloidal transport \cite{Lue12,Lue12a}. Thus one
would expect that the characteristic flow pattern due to the drift velocity
persists for small molecules. In a recent work, we have evaluated
thermodiffusion in binary mixtures in a mean-field model and found the Soret
coefficient \cite{Ler11}
\begin{equation}
S_{T}=\frac{\xi_{1}-\xi_{2}}{\phi_{1}D_{2}+\phi_{2}D_{1}}, \label{40}%
\end{equation}
where $\xi_{i}$ are thermodiffusion coefficients, $D_{i}=k_{B}T/6\pi\eta
R_{i}$ the tracer diffusion coefficients, and $\phi_{i}$ the volume fractions.
The denominator corresponds to the Hartley-Crank model for the mutual
diffusion coefficient \cite{Har56}. Its dependence on the molecular radii
$R_{i}$ constitutes a well-known hydrodynamic effect, that provides a good
description for quasi-ideal binary systems such as normal alkanes
\cite{Alo12}. The numerator of $S_{T}$ depends on the composition as
\[
\xi_{1}=\frac{2\beta}{9\pi\eta d_{0}}F_{1}(\phi_{1}H_{11}+\phi_{2}%
H_{12}),\ \ \ \ \ \ \xi_{2}=\frac{2\beta}{9\pi\eta d_{0}}F_{2}(\phi_{1}%
H_{21}+\phi_{2}H_{22}),
\]
where $\beta$ is the thermal expansion coefficient, $d_{0}$ a molecular
length, $H_{ij}$ usual Hamaker constants, and $F_{i}$ a correction factor that
depends on the molecular size. In the dilute limit $\phi_{1}\rightarrow0$, one
recovers an expression that describes experiments on polymer and particle
solutions \cite{Ler11}.

\subsection{Isotope effect}

Different isotopes of a given molecule differ in their thermodiffusion
coefficients, as shown by experiments \cite{Rut84,Deb01} and confirmed by
molecular dynamics simulations \cite{Mue99,Gal03,Art08}. Comparison with
kinetic theory for gas mixtures \cite{Wal49} suggests that this mass effect
arises from the kinetic energy of the molecules, with the mean value $\frac
{1}{2}m\overline{v^{2}}=\frac{3}{2}k_{B}T$. Since the mean momentum
$p\sim\sqrt{mk_{B}T}$ varies with the square root of mass and temperature,
lighter molecules and those coming from the hot side transfer more momentum
during a collision. By equilibrating this thermodynamic force with the Stokes
drag and imposing that there is no net force on a given volume element, it was
shown that lighter molecules are driven toward the warm because of their
stronger velocity fluctuations $\overline{v^{2}}$ \cite{Vil11}; the rotational
motion turns out to contribute significantly to the Soret coefficient,
agreement with experiment \cite{Deb01}.

This kinetic approach can be reformulated in terms of thermostatic quantities,
by noting that the molecular energy corresponds to the translational entropy
$S=k_{B}\ln V$ where $V$ is the available free volume per molecule. Explicit
formulae are given by Waldmann for ideal gases, in terms of the first virial
coefficients \cite{Wal49}. The volume $V$ is related to the molecular size and
the mean distance $\ell_{0}$ of Ref. \cite{Vil11}. Though these hard-sphere
models simplify the rather complex collisions of interacting molecules, they
agree well with molecular dynamics simulations of Lennard-Jones liquids
\cite{Mue99,Gal03,Art08}.

Artola et al. pointed out that the isotope effect is not accounted for by
thermostatic approaches such as Prigogine's model \cite{Art08}. This is
illustrated by a recent work claiming that the mass dependence of
thermodiffusion is a purely quantum phenomenon that vanishes when taking the
classical limit $\hbar\rightarrow0$ \cite{Har12}. On the other hand,
experiments in gases and liquids \cite{Wal49,Deb01}, numerical simulations
\cite{Art08}, and theory \cite{Vil11} concur to the conclusion that the mass
dependence of thermodiffusion is essentially a classical effect, implying that
the thermostatic approach of \cite{Har12} misses an important feature.

\section{Discussion and summary}

\subsection{Range of validity of Eastman's approach}

Eq. (\ref{18}) shows that for small particles the approach based on Onsager's
non-equilibrium thermodynamics reduces \ to a thermostatic expression for the
Soret coefficient. This has been traced back to the fact that drift and
diffusion of small particles lead to rather similar velocity fields
$\mathbf{v}_{T}$ and $\mathbf{v}_{D}$ \cite{Lon01,Mor08} and thus carry the
same friction factor. The latter drops out in the steady state, and $S_{T}$ is
determined by the solvation enthalpy $H$. For large particles, on the
contrary, $\mathbf{v}_{T}$ and $\mathbf{v}_{D}$ have little in common. Thus it
does not come as a surprise that the transport coefficients differ in their
viscous factors and, as a consequence, in their dependence on the particle
radius, $D\propto1/R$ and constant $D_{T}$. From a general transport theory
point of view, this analysis parallels what is known for electrophoresis,
where our Eqs. (\ref{22}) and (\ref{24}) correspond to the H\"{u}ckel and
Smoluchowski limits \cite{Hie97}.

These considerations are supported by experimental findings. Most data on
colloidal particles \cite{Put07,Bra08,Jia09,Esl12} confirm the linear
dependence $S_{T}\propto R$ of (\ref{28}), whereas Eastman's expression
(\ref{6}) predicts a variation with the surface area $S_{T}\propto R^{2}$. By
the same token, Eastman's equilibrium approach fails when applied to polymers.
The free enthalpy of a polymer chain increases with the number $N$ of monomers
and would result in $S_{T}\propto N$ according to (\ref{6}).\ Soret data on
dilute polystyrene solutions, however, vary with the gyration radius $R\propto
N^{\nu}$ where $\nu\approx0.6$ in a good solvent \cite{Gid76}, in agreement
with Brochard and de Gennes' general argument \cite{Bro81}, which is
explicited in our Eq. (\ref{28}). The law $S_{T}\propto R$ provides an
unambiguous signature for dissipative motion: The thermophoretic mobility
$D_{T}$ of polymers is constant because of the associated short-range viscous
flow, whereas the dependence $D\propto1/R$ arises from long-range hydrodynamic interactions.

\subsection{Binary mixtures}

The cases where the range of interaction $\lambda$ is much smaller or much
larger than the solute size $R$, can be treated in controlled approximations.
A more difficult situation arises for molecular mixtures with dispersion
forces, where both length scales are comparable. Though none of the
approximation schemes apply in this case, the relation $\lambda\sim R$
suggests that the thermostatic approch does not in general provide a reliable
expression for the Soret coefficient.\ One may expect that Eq. (\ref{40})
describes, at least qualitatively, the variation with the composition and the
molecular size ratio. Regarding the isotope effect, purely thermostatic
theories do not account for the mass dependence \cite{Art08}.

\subsection{Is Soret equilibrium an equilibrium effect?}

We have seen that for small solute particles, the Einstein and thermodiffusion
coefficients $D$ and $D_{T}$ carry the same mobility factor, such that the
ratio $D_{T}/D$ reduces to an equlibrium quantity. The physical origin of this
cancellation of viscous factors, however, is less fundamental than microscopic
reversibility that operates in Thomson's relation $\Pi/S=T$ for the Peltier
and Seebeck coefficients, or in the reciprocal law relating $D_{T}$ and the
Dufour coefficient \cite{Gro62}.

As a main result of this paper we have shown that a non-uniform solute
concentration in a temperature gradient implies necessarily dissipation due to
viscous flow $\mathbf{v}(\mathbf{r})$ in the vicinity of each solute particle,
as illustrated in Fig.\ \ref{Fig2}c. Diffusion and thermophoretic particle
currents in (\ref{8}) cancel each other, whereas the corresponding velocity
fields $\mathbf{v}_{T}+\mathbf{v}_{D}$ do not,\ thus resulting in a steady
entropy increase $\dot{S}>0$. Since any equilibrium state is characterized by
constant entropy, this means that the Soret coefficient in Eq. (\ref{2}) is
not an equilibrium quantity.

\end{document}